%% file: main.tex
\documentclass[10pt,conference]{IEEEtran}
\usepackage{amsmath,amssymb,amsfonts}
\usepackage{graphicx}
\usepackage{xcolor}
\usepackage{xspace}
\usepackage{url}
\usepackage{xurl}
\usepackage[normalem]{ulem}
\usepackage[hidelinks]{hyperref}
\usepackage[T1]{fontenc}
\usepackage[normalem]{ulem}

\usepackage{cite}

\usepackage{booktabs}
\usepackage{array}
\usepackage{tabularx}
\usepackage{multirow}
\usepackage{makecell}
\usepackage{colortbl}
\usepackage{threeparttable}
\usepackage{adjustbox}

\usepackage{subcaption}
\usepackage{svg}
\usepackage{tikz}
\usepackage{pgf}

\usepackage{enumitem}
\usepackage{pifont}
\usepackage{textcomp}
\usepackage{fontawesome}

\usepackage{listings}
\usepackage[most]{tcolorbox}
\usepackage{mdframed}

\usepackage{xparse}
\usepackage{ragged2e}
\usepackage{algorithmic}

\newcommand{\tool}[1]{{\small\textsf{#1}}}
\newcommand{\sysname}{\textsc{Harpoon}\xspace}
\newcommand{\siqi}[1]{{\color{blue}Siqi: #1}}
\newcommand{\liuhuo}[1]{{\color{pink}liuhuo: #1}}
\newcommand{\dk}[1]{\textbf{\color{green!50!black}dk: #1}}

\tcbuselibrary{listings,skins,breakable}

\lstdefinestyle{paperCode}{
  basicstyle=\ttfamily\scriptsize,
  numberstyle=\scriptsize\color{black!55},
  numbersep=4pt,
  breaklines=true,
  breakatwhitespace=false,
  columns=fullflexible,
  keepspaces=true,
  showstringspaces=false,
  tabsize=2,
  upquote=true,
  keywordstyle=\bfseries\color{blue!60!black},
  commentstyle=\color{gray!70!black},
  stringstyle=\color{teal!60!black},
  literate={+}{{\textcolor{green!55!black}{+}}}1
           {-}{{\textcolor{red!60!black}{-}}}1,
  emph={av_frame_unref,av_buffer_ref,av_buffer_unref,AVERROR,ENOMEM,decoder_ref,RefStruct,RKMPPDecoder,RKMPPFrameContext,CHECK_GE,LOG,include},
  emphstyle=\bfseries\color{purple!60!black},
  identifierstyle=\color{black!85},
}

\newtcolorbox[auto counter]{codebox}[2][]{%
  enhanced,
  breakable,
  colback=gray!3,
  colframe=black!15,
  boxrule=0.4pt,
  arc=2pt,
  left=8pt,right=6pt,top=6pt,bottom=6pt,
  title={Listing~\thetcbcounter:~#2},
  fonttitle=\bfseries,
  coltitle=black,
  attach title to upper,
  #1
}

\begin{document}


\title{Beyond a Single Commit: Anchor-Guided Reconstruction of Multi-Commit Patches}

\author{
\IEEEauthorblockN{
Jialiang Dong\IEEEauthorrefmark{1},
Willy Susilo\IEEEauthorrefmark{1},
Xu Gai\IEEEauthorrefmark{2},
Liuhuo Wan\IEEEauthorrefmark{1},
Chao Zhang\IEEEauthorrefmark{3},
Elisa Bertino\IEEEauthorrefmark{4},
and Siqi Ma\IEEEauthorrefmark{1}
}

\IEEEauthorblockA{
\IEEEauthorrefmark{1}
University of Wollongong
}

\IEEEauthorblockA{
\IEEEauthorrefmark{2}
University of International Relations
}

\IEEEauthorblockA{
\IEEEauthorrefmark{3}
Tsinghua University
}

\IEEEauthorblockA{
\IEEEauthorrefmark{4}
Purdue University
}
}

\maketitle

\begin{abstract}
\input{Tex/abstract}
\end{abstract}

\section{Introduction}
\input{Tex/intro-v2}

\section{Background}

\input{Tex/background}

\section{Methodology}
\input{Tex/methodology}

\section{Experiments}

\input{Tex/experiments}


\section{Related Work}
\input{Tex/relatedworks}

\section{Conclusion}
\input{Tex/conclusion}

\bibliographystyle{IEEEtran}
\bibliography{reference}

\end{document}

%% file: Tex/abstract.tex
Vulnerability patch identification plays a critical role in software security by linking reported vulnerabilities and the corresponding fixes,
    enabling reliable vulnerability analysis, mitigation, and automated security tool development. 
As software systems evolve in complexity, 
    vulnerability remediation has increasingly shifted from single-commit fixes to multi-commit fixing processes. 
Although many studies have been proposed to identify fixing commits associated with target vulnerabilities automatically, 
    they often suffer from inaccurate semantic analysis or excessive workload for language models. 
These limitations reduce both the effectiveness and comprehensiveness of fixing commit identification. 
    
To improve patch identification, 
    we propose an effective and comprehensive patch identification framework, \sysname, 
    which not only enhances the accuracy of identifying individual fixing commits, but also investigates dependencies among commits to achieve comprehensive commit recognition for target vulnerabilities. 
Experimental results show that \sysname significantly outperforms the state-of-the-art approaches. 
\sysname successfully identified 895 vulnerability patches that were not previously maintained in public vulnerability databases.
Among them, 663 are previously unrecorded fixes associated with 323 \texttt{CRITICAL}- or \texttt{HIGH}-severity vulnerabilities.

%% file: Tex/intro-v2.tex
With the widespread adoption and reuse of open-source software in modern software ecosystems, vulnerabilities in a single project can affect thousands or even millions of dependent applications. According to the 2025 OSSRA report~\cite{securitymag_ossra2025, synopsys_ossra2024}, 86\% of commercial codebases contain open-source vulnerabilities, while 81\% include \texttt{HIGH}- or \texttt{CRITICAL}-risk vulnerabilities.
Due to the large volume of the report vulnerability, 
    manually locating, verifying, and remediating all such vulnerabilities is often impractical for developers~\cite{first2026forecast}.
Hence,
    some vendors may employ vulnerability patch identification tools (e.g., Snyk~\cite{snyk}) to search for potential patch recommendations or rely on the patches maintained by the vulnerability repositories (e.g., NVD~\cite{nvd}, Debian~\cite{debian_Security}) to support remediation. 
Consequently, patch retrieval has become a fundamental task for supporting effective and efficient real-world vulnerability remediation, 
    requiring both high accuracy and comprehensive patch coverage. 


As modern software systems become increasingly complex, 
    such complexity has made vulnerability remediation an inherently iterative and ad-hoc process~\cite{nist80061}. 
This causes accurately and comprehensively identifying vulnerability patches challenging because \textbf{real-world vulnerability remediation often involves more than one fixing commit.} 
To recognize the fixing commits of a target vulnerability, 
    existing works commonly utilize three types of approaches --- semantic-based description analysis, large language model (LLM) reasoning, and hybrid analysis. 
Semantic-based description analysis~\cite{patchscout,vfcfinder,patchfinder,prospector} identifies fixing commits through semantic matching between vulnerability descriptions and code changes. 
However, real-world vulnerability descriptions and commit messages may be incomplete or inaccurate, 
    which can mislead semantic analysis by insufficient vulnerability information and superficial semantic similarities.
To enhance semantic understanding, 
    LLM-based reasoning~\cite{taper} has been proposed to leverage the reasoning capabilities of LLMs for analyzing commit messages and code changes. 
Nonetheless, 
    when numerous commits are involved in vulnerability remediation,
    LLMs struggle to process them effectively, leading to accumulated reasoning errors in large projects with long commit histories.
Although hybrid analysis~\cite{ship} is also proposed by combining semantic analysis with LLM reasoning, 
    it remains insufficient for overcoming the inherent limitations of semantic analysis and LLM reasoning.

To fill this gap,
    we propose an effective and comprehensive patch identification approach, \sysname, to reconstruct the patch for remediating a target vulnerability.
Given that a project repository may contain hundreds to millions of commits, and a single commit may involve multiple purposes (e.g., bug fixing, vulnerability remediation, or functionality changes~\cite{dong2025lies}),
    directly analyzing all commits is inefficient and prone to noise.
Therefore,
    \sysname first performs anchor commit filtering to identify \emph{anchor fixing commits}, namely, commits that are highly likely to be involved in remediating the target vulnerability.
For vulnerability remediation, two distinct scenarios are often involved: 1) similar vulnerabilities may be distributed across different locations in a project, requiring similar remediation logic;
2) complex vulnerabilities may require multiple fixing commits to fully resolve the underlying security issue,  
    \sysname accordingly employs two strategies --- \emph{coverage-completion} and \emph{dependency reconstruction}, to recover fixing commits that potentially contribute to the complete remediation process. 
Using the anchor fixing commits as the references, 
    \sysname then identifies candidate commits that share similar remediation patterns with the anchor commits or have direct/indirect dependency relationships with them.
Finally, \sysname performs an LLM-based refinement to understand the underlying intent of candidate commits and filter out those that exhibit different remediation intents despite sharing semantic similarities or dependency relationships with the anchor commits. 

We built a real-world vulnerability-patch benchmark dataset comprising 1,207 vulnerabilities. 
Of these, 1,019 were remediated by a single fixing commit (denoted as \emph{single-commit patches}), while 188 required multiple fixing commits over different time periods (denoted as \emph{multi-commit patches}).
We assessed the effectiveness and comprehensiveness of \sysname in identifying patches by applying it to the vulnerability-patch benchmark and comparing its performance against four state-of-the-art approaches. 
The results show that \sysname substantially outperforms state-of-the-art approaches for both single-commit and multi-commit patch identification, 
    achieving an F1 of 93.02\% and 75.08\%, respectively, 
    while the best state-of-the-art approach, \tool{Prospector}, only achieves 90.85\% and 55.21\%, respectively. 
\sysname also demonstrates strong efficiency, reducing the median processing time from 130 seconds achieved by the state-of-the-art approach to only 40 seconds.
We further applied \sysname to an in-the-wild dataset and identified 323 previously undocumented patches for \texttt{CRITICAL}- or \texttt{HIGH}-severity vulnerabilities,
    which were subsequently reported to CVE, Debian Security, and GHSA.
Our evaluation shows that 44.58\% of these vulnerabilities require multi-commit patches for complete remediation, 
    indicating that the identification of a single fixing commit is insufficient for recovering the vulnerability remediation process. 
Among the identified multi-commit patches,
    nearly half (48\%) are completed within one week, whereas 13\% take longer than one month to complete.

\noindent
\textbf{Main contributions:}

\begin{itemize}[leftmargin=1.2em]
\item \textbf{An effective and comprehensive patch reconstruction approach. } We designed two patch reconstruction strategies to address the mainstream vulnerability remediation practices. 
The strategies support the comprehensive identification of both single-commit and multi-commit patches, 
    while also revealing the correlations among the fixing commits within each multi-commit patch. 
This allows developers to understand clearly how fixes work together to achieve complete remediation of the target vulnerability.

\item \textbf{A lightweight commit identification tool.} We built \sysname, to focus on fixing commits that are highly likely to be associated with the target vulnerability, 
    avoiding exhaustive analysis of all commits in the repository. 
This reduces the noise introduced during semantic analysis and the computational cost of commit analysis.  



    \item \textbf{Large-scale empirical findings.} Using \sysname, we uncover 323 previously undocumented high-severity vulnerability patches, including 144 multi-commit patches. 

\end{itemize}

\noindent \textbf{Ethics and Disclosure.} 
To uphold ethical research practices, 
all detected previously undocumented multi-commit patches were reported to CVE, Debian Security, and GHSA.
We release our dataset and \sysname implementation to facilitate future research~\cite{artifact}. 

%% file: Tex/background.tex
\subsection{Multi-Commit Vulnerability Remediation}

    

Following the vulnerability disclosure policy, 
    software vendors are expected to fix affected products either prior to disclosure or immediately after the vulnerability is publicly disclosed.
Additionally,   
    real-world practices show that vulnerability remediation does not always proceed in a linear ``analyze once, fix once, release once'' manner. 
Developers may need to identify additional instances of the same vulnerability type across the project, 
    or continuously refine the remediation to address follow-up issues introduced by earlier patches~\cite{iso_iec_30111}.
We refer such cases as \textbf{multi-commit patches}.
    In contrast, cases involving only a single fixing commit are called \textbf{single-commit patches}.

To identify vulnerability-fixing commits, existing approaches mainly rely on three representative techniques: semantic-based vulnerability description analysis~\cite{patchscout,vfcfinder,patchfinder,prospector}, LLM-based reasoning~\cite{taper}, and hybrid methods~\cite{ship}:

\noindent
\textbf{Semantic-based} analysis focuses on retrieving candidate commits by matching vulnerability descriptions with code changes or commit messages. However, they assume that semantic similarity implies remediation relevance, making them susceptible to incomplete vulnerability descriptions and semantically similar but unrelated commits. 


\noindent
\textbf{LLM-based} reasoning
leverages the semantic analysis and reasoning capabilities of LLMs to determine whether a candidate commit contributes to remediating a target vulnerability.
Although such approaches improve semantic understanding of fixing commits, 
    the LLM capabilities are commonly inconsistent and unreliable, 
    especially when numerous fixing commits are involved. 
    

\noindent
\textbf{Hybrid approach} combines semantic analysis with LLM reasoning. However, they identify additional commits primarily through commit similarity and do not explicitly model progressive remediation behaviors, limiting their ability to recover real-world multi-commit patches.

\subsection{Motivating Example and Challenges}
\begin{figure}[t]
\centering

\begin{subfigure}{\linewidth}
    \centering
    \includegraphics[width=0.95\linewidth]{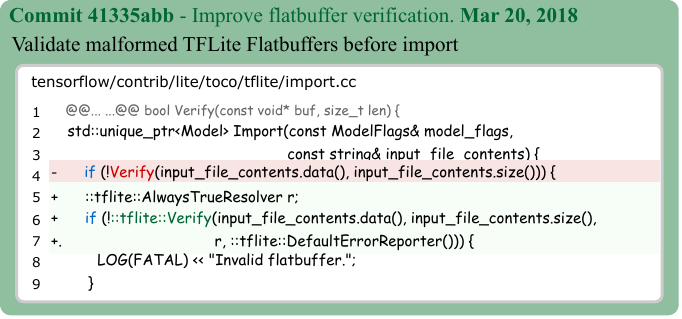}
    \caption{Fixing commit~\texttt{41335abb}.}
    \label{fig:motivating_case_fix1}
\end{subfigure}

\vspace{0.5em}

\begin{subfigure}{\linewidth}
    \centering
    \includegraphics[width=0.95\linewidth]{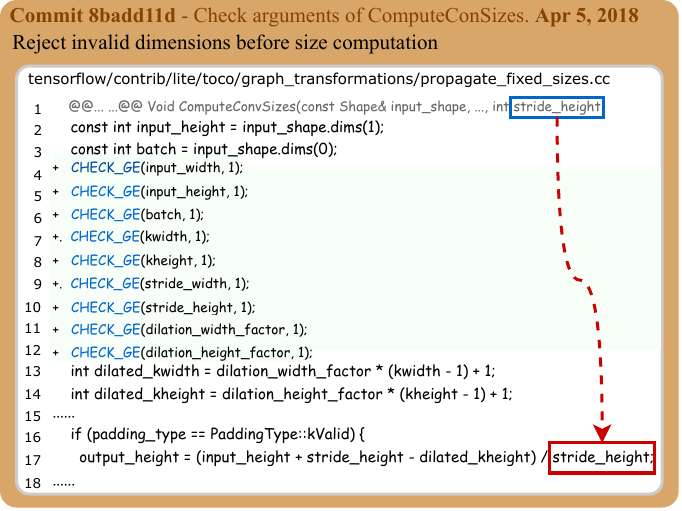}
    \caption{Fixing commit~\texttt{8badd11d}.}
    \label{fig:motivating_case_fix2}
\end{subfigure}

\caption{
Motivating Example based on CVE-2018-8825.
}
\label{fig:motivating_example}

\end{figure}

Figure~\ref{fig:motivating_example} illustrates a motivating example of a vulnerability remediated through a multi-commit patch. This case cannot be identified effectively by existing approaches. 
Vulnerability \textbf{CVE-2018-8825~\cite{CVE-2018-8825}} is a \texttt{HIGH}-severity buffer overflow vulnerability (CVSS 8.8), 
    which can be triggered by a crafted TFLite model and may result in program crashes or arbitrary code execution.
According to the publicly available vulnerability information,
    no patch information is maintained by vulnerability repositories.
Hence,
    the corresponding remediation commits are not immediately available and must be inferred from additional project artifacts and repository histories.

However, 
    when we manually check the fixing commits given by TensorFlow in its repository, 
    it first addressed the vulnerability in commit \texttt{41335abb} on March 20, 2018 (shown in Figure~\ref{fig:motivating_example}(\subref{fig:motivating_case_fix1})).
This commit replaced the flawed \texttt{Verify} function at Line 4 with stricter validation logic (Lines 5--7), ensuring that malformed TFLite flatbuffer inputs are properly checked for structural validity.
Unfortunately, 
    such structural validation is insufficient to fully mitigate this vulnerability because structurally valid models may still contain \textbf{semantically} invalid convolution parameters (e.g., zero dimensions).
These invalid parameters can trigger erroneous shape inference or unsafe memory access during subsequent computation.
TensorFlow deverlopers further recognized such an issue and introduced the second commit \texttt{8badd11d} on April 5, 2018 (shown in Figure~\ref{fig:motivating_example}(\subref{fig:motivating_case_fix2})), 
    which added validity checks for parameters such as \texttt{stride\_height} to ensure semantic validation (Lines 9, 17).

When applying existing approaches, commit \texttt{41335abb} can be readily identified as the fixing commit of CVE-2018-8825
because the changes directly processed with the disclosed vulnerability. 
However, 
    it is challenging to identify the follow-up commit \texttt{8badd11d} because:
\begin{enumerate}
\item The CVE-2018-8825 report only contains a coarse-grained description of the vulnerability impact and triggering conditions. 
It only states that TensorFlow is affected by a buffer overflow and contains no indication of the semantic validation issue addressed by \texttt{8badd11d}, limiting semantic-based approaches (\textit{Challenge \#1}).

\item Commits~\texttt{41335abb} and \texttt{8badd11d} indicate a progressive remediation process distributed across different files and code locations, separated by more than two weeks. Existing approaches typically analyze commits in isolation, making them unable to capture remediation dependencies between related commits (\textit{Challenge \#2}).

\item TensorFlow introduced 586 other commits between Commits~\texttt{41335abb} and \texttt{8badd11d}. 
These intervening irrelevant commits substantially increase the analysis noise,
    and make it difficult for existing approaches to identify the relevant remediation commits.  (\textit{Challenge \#3}).

\end{enumerate}

%% file: Tex/methodology.tex
\begin{figure*}[t]
  \centering
  \includegraphics[width=0.9\textwidth]{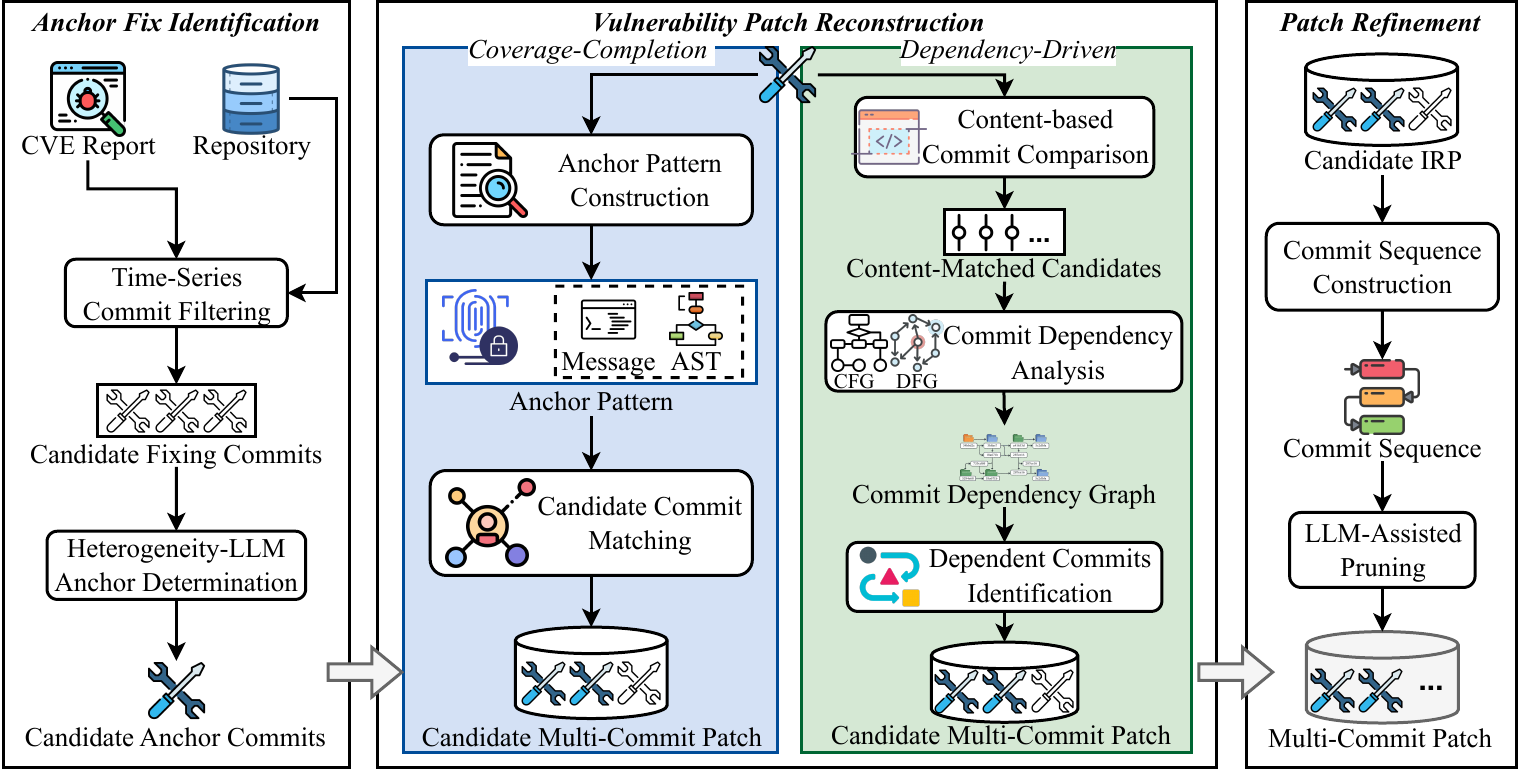}
  \caption{Overview of \sysname.}
  \label{fig:overview}
\end{figure*}

To identify the patches of the target vulnerability comprehensively,
    we propose a framework \sysname, shown in Figure~\ref{fig:overview}.

\subsection{Anchor Fix Identification} 
\label{subsec:anchor_fix}
Instead of directly analyzing a large number of fixing commits, 
    \sysname leverages commit features to identify anchor commits for each target vulnerability.
According to the vulnerability fix pattern --- fixes associated with a vulnerability are usually committed within a limited time window close to the vulnerability public disclosure date~\cite{li2017large}, 
    \sysname first conducts time-series commit filtering to narrow down the scope of commit analysis. 
Then it uses vulnerability features to identify the commits with vulnerability-related features through heterogeneity-LLM anchor determination.

    


\subsubsection{Time-Series Commit Filtering}
To determine the time-series range,
    \sysname extracts the CVE disclosure date and examines the affected software versions listed in the \textit{Known Affected Software Configurations} field. 
Specifically, 
    it conducts the initial filtering by selecting fixing commits made between the release dates of the last affected version and that of the first fixed version as potential fixing commits associated with the target vulnerability.
When the affected software version is unavailable in the CVE report, 
    \sysname applies a severity-dependent look-back window from the CVE publication date~\cite{cisaRef,necleussecBlog,xmcyber}, i.e., 30 days before publication for \texttt{CRITICAL}/\texttt{HIGH} vulnerabilities,
60 days before publication for \texttt{MEDIUM} vulnerabilities,
and 90 days before publication for \texttt{LOW} vulnerabilities.
If the initial filtering cannot locate any fixing commits,
    \sysname further extends the search range to six months before/after the CVE publication date.

\subsubsection{Heterogeneity-LLM Anchor Determination}
Since LLMs have an intelligent performance in understanding the semantic meaning of text, 
    \sysname employs LLM-based dependency analysis to examine correlations among the candidate commits for anchor commit identification.
However, 
    even after narrowing down the commit scope through time-series filtering, the number of candidate fixing commits may remain large, 
    making direct LLM-based commit processing infeasible.
    
Therefore, 
    \sysname first filters fixing commits using the vulnerability features because commits containing richer vulnerability-related information are more likely to be relevant to the target vulnerability.
It then applies an LLM to analyze the commit descriptions and determine the anchor commits.

\noindent
\textbf{Heterogenity-aware selection.}
Vulnerability features have different levels of uniqueness - higher-uniqueness features are more informative for identifying relevant fixing commits than lower-uniqueness features.
To achieve comprehensive coverage, 
    we examine four types of vulnerability features in descending order of uniqueness:
    Vulnerability Identifiers $>$ Commit Hashes $>$ Vulnerable Elements $>$ CWE Keywords.     




\begin{itemize}[leftmargin=1.2em]
\item \textbf{Vulnerability Identifiers:} These include \texttt{CVE-ID}s and related issue IDs, which can be used in vulnerability reports to uniquely distinguish individual vulnerabilities.


\item \textbf{Commit Hashes:} 
Commit hashes referenced by vulnerability reports to link vulnerabilities with their corresponding fixing commits.


\item \textbf{Vulnerable Elements:} These are the code-related information that might be highlighted in CVE reports, including affected files, functions, and variables.

\item \textbf{CWE Keywords:} These are the keywords for describing the vulnerability type, which include \textit{Name}, \textit{Description}, \textit{Alternate Terms}, \textit{Weakness Ordinalities}, \textit{Potential Mitigations}, and \textit{Demonstrative Examples}. 

\end{itemize}



\sysname first extracts the vulnerability features using regex matching with tailored patterns.  
For retrieving vulnerability identifiers, 
    \sysname uses the pattern \texttt{CVE-[0-9]\{4\}-[0-9]+} to analyze the CVE \texttt{References} field, 
    and checks the issue references through patterns \texttt{/issues/[0-9]+} or \texttt{/pull/[0-9]+}. 
In terms of commit hashes, 
    \sysname matches URLs containing both \texttt{git} and \texttt{commit},
    and further extracts the corresponding commit hashes using the regex pattern
    \texttt{[0-9a-f]\{40\}}.
Since vulnerable elements are often expressed as natural-language phrases, 
    it makes them difficult to distinguish from the original vulnerability descriptions. 
Hence,
    \sysname employs \texttt{GPT-5-Nano}~\cite{openai_gpt5_nano}, 
    which exhibit strong capabilities in natural-language understanding, 
    to analyze the CVE description and extract vulnerable code elements, including affected files, functions, and variables.
To collect CWE keywords, 
    \sysname first builds a vulnerability keyword corpus from the content of each CWE type.
Then it applies the Term Frequency-Inverse Document Frequency (TF-IDF) algorithm to estimate term importance and selects the top-3 terms for each CWE as its representative keywords.
Given the vulnerability keyword corpus, 
    \sysname matches the keywords of each vulnerability type against the CVE report. 
Once any keyword is matched, 
    the CVE report is grouped into the corresponding vulnerability type, 
    and all keywords associated with that type are used as the representative keywords of the report for subsequent fix analysis.

Having the vulnerability features, 
    \sysname performs feature matching over commit messages and code diffs following their uniqueness order.
Once a commit contains a feature type with higher uniqueness, 
    \sysname terminates the matching process and skips the comparison with the remaining lower-uniqueness features.

\noindent
\textbf{LLM-based commit analysis.}
To better retrieve the correlations among the commits, 
    \sysname utilizes \texttt{GPT-5-Nano} to analyze the commits that contain vulnerability features, using their commit messages and code diffs, along with the vulnerability description, as the input.
\sysname finally outputs the commits identified as genuine vulnerability fixing commits as anchor commits.

\subsection{Vulnerability Patch Reconstruction}
A vulnerability may be fixed through multiple commits because 
    1) the same vulnerability can affect multiple code locations requiring similar remediation logic;
    2) the vulnerability is difficult to fix comprehensively with a single patch, or earlier patches may introduce some subsequent bugs that require additional fixes.  
Based on the two cases, 
    \sysname analyzes fixing commits by employing two complementary reconstruction strategies, \textit{Coverage-Completion} and \textit{Dependency-Driven}, to analyze relationships between anchor commits and remaining fixing commits, thereby identifying potential multi-commit patches.
    
When fixing a vulnerability in practice, 
    developers often inspect not only the specific code location where the vulnerability is identified, 
    but also the entire project to determine whether similar vulnerabilities are present in other locations.
If so,
    developers may apply similar fixes to those vulnerable code snippets.
It indicates that similar fixes for the same vulnerability are typically committed within a certain period around the publication date of the anchor commit.
Accordingly,
    \sysname restricts the analysis scope to commits published within 33 days before and after the anchor commit~\cite{li2017large} for patch reconstruction, 
    rather than analyzing all fixing commits.

\label{subsec:irp_recons}

\subsubsection{Coverage-Completion Patch Reconstruction}
This strategy addresses the scenario where the same vulnerability spans multiple code locations (e.g., different files or modules), 
    and requires similar fix logic across these locations.
\sysname leverages the patterns of anchor commits as reference to identify other commits with similar code-change logic and remediation-description patterns.



\noindent
\textbf{Anchor Pattern Construction.}
From each anchor commit,  
    \sysname separately extracts two types of patterns from two objectives: code changes and commit descriptions. 
For code-change analysis, 
    \sysname first decomposes code diffs and generates hunks where each hunk represents a contiguous block of modified code lines surrounding a specific change.
Within each hunk, 
    \sysname then parses the pre-fix and post-fix code fragments into ASTs using \tool{tree-sitter}~\cite{tree_sitter}.
It further identifies AST-delta features by comparing node-type frequency differences between the pre-fix and post-fix ASTs, 
    thereby capturing the structural code changes introduced by the fix.
To extract the remediation description patterns, 
    \sysname analyzes commit descriptions and constructs a TF-IDF model over commit messages. 
It collects the highest-weighted terms from the descriptions.

\sysname constructs a fix fingerprint for each anchor commit by combining the patterns of the code changes and commit descriptions.

\noindent
\textbf{Candidate Commit Matching.}
Taken as input the fix fingerprint of each anchor commit, 
    \sysname further compares the fix fingerprint with the other commits to identify those containing similar fix patterns.
Specifically,
    \sysname applies the same AST-delta feature extraction and TF-IDF encoding procedure to each target commit as described for the anchor commit.
It then conducts hunk-level comparison by calculating the cosine similarities between the AST-delta TF-IDF vectors of the anchor commit and the target commit.
According to the similarity scores,
    \sysname first selects the top-$k$ most similar commits ($k=10$ by default) based on code-change similarity, and then verifies whether their commit descriptions contain the highest-weighted terms extracted from the anchor commit message.
Commits satisfying both the code-change similarity and commit-description consistency criteria are retained as the coverage-completion candidate set for the target multi-commit patch.

\subsubsection{Dependency-Driven Patch Reconstruction}
This strategy targets vulnerabilities whose remediation is distributed across multiple dependent fixing commits.
Such commits may correct issues introduced by earlier patches or replace incomplete fixes.
\sysname identifies commits correlated with the anchor commit through forward and backward slicing.


\noindent
\textbf{Content-based Commit Comparison.}
\sysname selects the candidate commits that share similar elements with the anchor commits.
Specifically, \sysname collects three levels of comparison between the anchor commits and the candidate commits: file, code, and metadata.
It first applies file-level comparison to examine similarities in modified directories and file names. 
Then,
    it conducts code-level comparison by comparing modified functions and variables. 
Finally, 
    \sysname conducts metadata-level comparison to match author information, issue IDs, pull-request IDs, bug IDs, and commit hashes from the anchor commit.
While most metadata fields can be extracted directly,
    bug IDs are often referenced in customized formats.
To address this issue,
    \sysname first locates potential bug-ID mentions using indicative terms, including \textit{follow-up}, \textit{fixup}, \textit{regression}, \textit{tighten}, \textit{complete}, \textit{completed}, \textit{complement}, and \textit{harden},
    and then extracts bug IDs through regex matching.

\sysname builds a commit candidate $\mathcal{S}_1(a)$, 
    where $a$ represents the anchor commit used for comparison. 
When the target commit $c$ matches any one element with the anchor commit $a$,
    \sysname adds it to $\mathcal{S}_1(a)$.

\noindent
\textbf{Commit Dependency Analysis.}\label{sec:commit-dep}
Building upon $\mathcal{S}_1(a)$,
    \sysname further analyzes whether the modifications introduced by candidate commits exhibit program-semantic dependencies with those introduced by the anchor commit.
Specifically,
    \sysname retrieves the contextual semantics of both the anchor commit and candidate commits, and determines whether their modifications are dependent.
    
Given each anchor commit $a$, 
    \sysname first analyzes its pre-fix and post-fix versions to extract the affected functions, 
    and then uses \tool{tree-sitter}~\cite{tree_sitter} to parse the corresponding functions into ASTs. 
To better capture data/control dependencies, 
    \sysname separately converts the ASTs into control-flow graphs (CFGs) and data-flow graphs (DFGs).
For CFG construction, 
    \sysname maps AST nodes corresponding to execution statements, 
    including assignment statements, conditional statements, function calls, and return statements,
    into graph nodes.
The control-flow edges are then added between two nodes based on the code execution orders.
For DFG construction, 
    \sysname starts from variable declarations in the AST to identify the variables that may be involved in vulnerability remediation. 
It then traces the data flow of each variable by analyzing how it is used throughout execution. 
The involved variables are represented as graph nodes, 
    and directed edges are added between nodes to represent data flow execution.

To retrieve the nodes that potentially contribute to remediating the target vulnerability, 
    \sysname first compares the CFGs/DFGs constructed from the pre-fix and post-fix versions and separately identifies the added and deleted nodes. 
It then starts from the changed nodes and performs forward and backward traversals over CFGs and DFGs to collect additional nodes with direct/indirect control/data dependencies. 
All dependent nodes,
    together with the original nodes, are included in a node set $N(a)$. 
For each candidate commit $c\in S_1(a)$,
    \sysname performs the same step to construct a node set $N(c)$. 

To further extract the correlations between $N(a)$ and $N(c)$, 
    \sysname compares the nodes in $N(a)$ and $N(c)$ pairwisely. 
When the same node exists in both $N(a)$ and $N(c)$, 
    it adds a directed edge between the corresponding node in the dependency graphs of the candidate commit and the anchor commit. 
According to the connected graphs between the candidate commits and each anchor commit, 
    \sysname selects the fixing commits that appear in both $S_1(a)$ and the connected graph.
These commits are then recovered and stored together with the anchor commit as potential fixing commits belonging to the multi-commit patch of the target vulnerability, 
    while preserving their dependency relationships.

\subsection{Multi-Commit Patch Refinement} \label{subsec:irp_refine}
Although \sysname retrieves potential fixing commits that are likely to contribute to the same vulnerability remediation, 
    the candidate set may still include commits that modify the same elements as the anchor commits but serve different purposes. 
For example, 
    commits that only update \texttt{print} outputs may be included despite being unrelated to vulnerability remediation.
\sysname thus applies LLM-based semantic refinement to investigate the intent behind modification of each commit.


In detail, 
    \sysname organizes the candidate commits of a multi-commit patch into a chronological sequence based on their commit timestamps.
Then \sysname formulates task-oriented prompts by including the commit-related information (i.e., commit hash, timestamp, commit description, and code diff), 
    together with the associated vulnerability description, 
    to guide the LLM to confirm the fixing commits through  semantic-level intent analysis. 
The complete prompt is available in our artifact repository~\cite{artifact}.

%% file: Tex/experiments.tex
\subsection{Experimental Settings}
To evaluate the effectiveness and comprehensiveness of \sysname in identifying patches for target vulnerabilities, 
    we answer the following four research questions (RQs):

\begin{itemize}[leftmargin=1.2em]
    \item \textbf{RQ1: Identification Effectiveness.} How accurately can \sysname identify patches of the vulnerability?
    
    \item \textbf{RQ2: Ablation Study.} How does each component of \sysname contribute to its overall performance?


    \item \textbf{RQ3: Computational Cost.} What is the computational overhead of \sysname?

    \item \textbf{RQ4: In-the-Wild Evaluation.} Can \sysname effectively identify patches in real-world programs?

\end{itemize}


\subsubsection{Dataset}
Since there is no existing dataset providing such multi-commit patch for each CVE,  
    we first construct a vulnerability-patch benchmark dataset with comprehensive patch annotations. 
    
\noindent
\textbf{Data collection.}
Due to the impracticality of analyzing all CVEs, 
    we selected the top-10 popular open-source projects, 
    written in C/C++, Java, and Python, respectively, each with more than 1K GitHub stars.
In total, 
    we retrieved 30 projects. 
We then collected all CVE-listed vulnerabilities associated with these projects. 
Because fixing commits are not consistently maintained on CVE pages, 
    we supplemented the fixing commit information for each CVE by manually retrieving relevant commits from multiple sources, including NVD~\cite{nvd}, Debian Security~\cite{debian_Security}, and the GitHub Advisory Database~\cite{ghsa}.
Eventually, 
    we obtained 2,374 CVEs with publicly available fixing commits, of which 562 involved multi-commit patches. 

\noindent
\textbf{Benchmark construction.}
Although we collected fixing commits from various sources, 
    some patches remain missing from publicly maintained vulnerability databases~\cite{dunlap2024vfcfinder}.
In addition, 
    the definitions of multi-commit patches across different vulnerability management platforms are slightly different.
For instance, 
    GitHub Advisory Database, NVD, and the benchmark given by SHIP~\cite{ship}  treat identical fixes applied to different locations within the same project as multi-commit patches;
    contrarily, we treat them as single-commit patches.
Hence, 
   automatically retrieved vulnerability-patch data may be incomplete or inconsistent, 
   making it unsuitable for direct use as a benchmark ground truth for evaluating the effectiveness and comprehensiveness of patch identification in the paper.

Therefore,
    we manually inspected all 562 vulnerabilities recorded with multi-commit patches and a random sample of 1,200 vulnerabilities recorded with single-commit patches. 
Specifically,
    two PhDs with over five years of software security research experience independently checke and refined the collected data using the following procedures:


\begin{itemize}[leftmargin=1.2em]
\item \textbf{Backport Deduplication.}
The same patch may be applied to multiple software versions, 
    causing backported commits to be mistakenly counted as distinct patches during automated commit collection.
Following the approach proposed by Shariffdeen et al.~\cite{shariffdeen2021automated},
    we analyzed the semantics of each fixing commit to determine whether any semantically-equivalent commits existed in any previous versions of the project.


\item \textbf{Remediation Chain Completion.}
Publicly maintained patches are often incomplete --- 
    some vulnerabilities may only have a small subset of their associated patches available.
We further manually examined fixing commits published within a temporal window around each collected fixing commit. 
Based on existing observations~\cite{li2017large} that related patches are typically committed within 33 days of one another, we thus set the search window to 33 days before and after the commit date of each collected fixing commit.
If multiple fixing commits were detected, we iteratively searched for additional nearby fixing commits until the remediation chain can no longer be expanded.


\end{itemize}



They then cross-validated the results to analyze the disagreements. 
Disagreements were resolved through discussion to reach a consensus with a Kappa score = 0.86, indicating an almost perfect level of agreement~\cite{landis1977measurement}.
We finally constructed a \emph{vulnerability-patch benchmark} consisting of patches for 1,207 vulnerabilities. 
These include 188 multi-commit patches and 1,019 single-commit patches, comprising 1,583 fixing commits. 

\subsubsection{Default settings}
\sysname employs \texttt{GPT-5-Nano}~\cite{openai_gpt5_nano} as the default LLM due to its favorable balance between effectiveness and computational cost.

\subsection{RQ1: Identification Effectiveness}
\label{subsec:rq1}
We applied \sysname to the vulnerability-patch benchmark to evaluate its effectiveness in identifying patches for each target vulnerability and its comprehensiveness in covering all fixing commits associated with the vulnerability. 
Furthermore, 
    we compared \sysname against four state-of-the-art patch identification approaches, \tool{TAPER}~\cite{taper}, \tool{VFCFinder}~\cite{vfcfinder}, \tool{SHIP}~\cite{ship} and \tool{Prospector}~\cite{prospector}, 
    as they have been shown to outperform other existing approaches.

\subsubsection{Experiment Setup}
The definition of multi-commit patch identification varies slightly across existing studies, 
    and most patch identification approaches focus on identifying single-commit patches, which makes the comparison challenging. 
Correspondingly,
    we conducted three types of comparisons to separately assess commit identification effectiveness and commit coverage comprehensiveness:


\begin{itemize}
\item \textbf{Commit-Level comparison (abstract as \texttt{Commit}).} This evaluates commit identification effectiveness. 
A fixing commit is correctly identified if it belongs to the ground-truth patch of the target vulnerability. 

\item \textbf{CVE-Level Partial Matching (\texttt{CVE-P}):} This evaluates the effectiveness of identifying vulnerability-specific patches.
An identified patch is labeled correct if every fixing commit within the patch is included in the ground-truth patch of the target vulnerability, without introducing any irrelevant commits.

\item \textbf{CVE-Level Exact Matching (\texttt{CVE-E)}:} This assesses the comprehensiveness and effectiveness of identifying vulnerability-specific patches. 
An identified patch is considered correct only if it includes all commits in the ground-truth patch of the vulnerability and does not introduce any irrelevant commits.

\end{itemize}

Notably, 
    \tool{VFCFinder} and \tool{Prospector} formulate vulnerability-specific patch identification as a ranking task, 
    where they return a ranked top-$r$ candidate commits instead of directly outputting a single-commit or a multi-commit patch. 
Without knowing whether a target vulnerability involves a single-commit or a multi-commit patch, 
    we considered two settings for determining the value of $r$:
    1) we used the number of fixing commits identified by \sysname as the hint of the commit number associated with the vulnerability, and set $r$ accordingly (denoted as \tool{Tool}$_\textsc{Hint}$);
    2) we followed the top-ranked evaluation mentioned in the original papers to set $r$ accordingly, i.e., $r=5$ in \tool{VFCFinder}~\cite{vfcfinder} and $r=10$ in \tool{Prospector} (denoted as \tool{Tool}$_\textsc{Orig}$).

\input{Tables/rq1_srp}

\input{Tables/rq1_irp_results}

\subsubsection{Results Analysis}
Table~\ref{tab:rq1_srp} and Table~\ref{tab:irp_benchmark} present all the fix identification results for single-commit patches and multi-commit patches, respectively.
Among 1,207 vulnerabilities,
    \sysname successfully identified single-commit patches for 886 out of 1,019 vulnerabilities and multi-commit patches for 104 out of 188 vulnerabilities, 
    outperforming all state-of-the-art approaches.

\noindent
\textbf{Vulnerabilities with single-commit patch.} 
\sysname achieves an commit identification precision of 92.87\%, a recall of 93.17\%, and an F1 of 93.02\%.
Although \tool{Prospector} correctly identified single-commit patches for 848 vulnerabilities, achieving an F1 of 90.84\%,
    it introduced 100 false positives, causing a lower precision (89.45\%), ultimately reducing the entire performance.
\tool{VFCFinder} and \tool{SHIP} only correctly identified 76 and 603 vulnerabilities with single-commit patches, respectively. 

After manually inspecting the failed cases, 
    we found that \tool{VFCFinder}, \tool{SHIP}, and \tool{Prospector} usually misclassify functional-improvement and refactoring commits as vulnerability-specific fixing commits. 
As these approaches rely heavily on lexical matching and semantic similarity between vulnerability descriptions and commit information,
    they fail to differentiate fixing commits from semantically similar but irrelevant commits.
\tool{Taper} performs the worst, 
    failing to identify any correct fixing commits because it heavily relies on LLMs for fixing-commit identification.
However, 
    LLMs struggle to accurately identify fixing commits when a large number of candidate commits are provided, especially when many content-similar, but vulnerability-irrelevant commits appear. 

Instead, 
    \sysname combines the patterns of both code modifications and commit descriptions for patch identification, which are less affected by such misleading textual similarities. 
We also manually inspected the 133 cases misclassified by \sysname and observed 64 cases are caused by insufficient vulnerability-reference information. 
Although \sysname supports vulnerability analysis at different levels of granularity, 
    ranging from fine-grained vulnerability identifiers to coarse-grained vulnerability keywords,
    some CVE reports remain too vague and contain only a single-sentence description (e.g., ``Ansible prior to 1.5.4 mishandles the evaluation of some strings''~\cite{CVE-2014-2686}).
Hence,
    \sysname failed to pinpoint anchor commits effectively for subsequent fixing commit identification.
In addition, 
    62 cases failed because some vulnerability types share similar CVE features and code-change patterns. 
For example, 
    buffer overflows and input-validation errors may involve similar vulnerable inputs and boundary checks. 
When only coarse vulnerability features are available, \sysname is difficult to accurately distinguish patches for similar vulnerability types.
7 cases were failed because of the large time gap between the fixing commits and the vulnerability disclosure date,
    which makes our predefined temporal window unable to cover the fixing commits.

\noindent
\textbf{Vulnerabilities with multi-commit patches.}
Multi-commit patch identification requires recovering a sequence of fixing commits,
    demanding both effective commit detection and comprehensive commit coverage. 
The results demonstrate that \sysname outperforms the state-of-the-art tools in terms of both effectiveness and comprehensiveness across all settings, 
    including \textbf{\texttt{Hint}} and \textbf{\texttt{Orig}} for ranking-based approaches. 

While considering effectiveness (i.e., \textbf{\texttt{Commit}}), 
    \sysname successfully identified 355 out of the 564 fixing commits, achieving 93.13\% precision, 62.89\% recall, and an F1 score of 75.08\%, substantially outperforming all baselines. 
Among the state-of-the-art approaches,
    \tool{Prospector}$_\textsc{Hint}$ achieved the strongest baseline performance, with a precision of 74.91\%, a recall of 43.70\%, and an F1 score of 55.21\%.
However,
    its recall remained considerably lower than that of \sysname because \tool{Prospector} ranked the fixing commits independently without considering their correlations. 
Consequently, 
    some actual fixing commits with low textual similarity to the vulnerability description, despite being related to the same remediation process, received lower rankings.
\tool{VFCFinder} also failed to retrieve the commit correlations because of the similar ranking mechanism.   
Although \tool{SHIP} examined relationships among commits, 
    its analysis depends heavily on the quality of vulnerability reports and only performed well on high-quality reports. 
\tool{Taper} is heavily restricted by the effectiveness of LLMs, 
    which performed the worst.

\input{Tables/rq2_ablation_study}

Regarding the comprehensiveness commit coverage (i.e., \textbf{\texttt{CVE-E}}
and \textbf{\texttt{CVE-P}}), 
    \sysname performs consistently well, which exactly identified 104 out of 188 patches with 63.8\% precision, 80.62\% recall, and an F1 of 71.23\%, 
    and partially identified 138 of them, yielding 84.66\% precision, 84.66\% recall, and an F1 of 84.66\%.
By comparison, 
    existing approaches suffer from a significant performance degradation when comprehensiveness is considered, 
    even under partial matching.
It indicates that existing solutions struggle to deal with multi-commit patch identification. 

Through our manual observation of the failure cases of \sysname, 
    we found that the anchor commits for 24 vulnerabilities were incorrectly identified due to insufficient vulnerability-reference information and erroneous LLM-based extraction of vulnerable code elements.
The limitation is discussed further in Section~\ref{sec:discussion}.
In terms of the missing commits for 34 vulnerabilities,
    19 cases were missed because their remediation patterns were uncommon and therefore not covered by any identified anchor commit;
    the remaining 15 failed to be recognized by the current \texttt{tree-sitter}-based dependency analysis.

\subsection{RQ2: Ablation Study}
We further assessed the contribution of each module to the patch identification. 
We separately removed the \textit{Coverage-Completion} reconstruction module (w/o \textit{Coverage}), the \textit{Dependency-Driven} reconstruction module (w/o \textit{Dependency}), and the \textit{Patch Refinement} module (w/o \textit{Refinement}), and measured the performance changes. 
The experimental results are presented in Table~\ref{tab:component}.

\noindent
\textbf{Coverage-Completion Patch Reconstruction.}
When \textit{Coverage-Completion} was removed, 
    the results of correct commit identification remained almost unchanged.
However,
    \sysname overlooked 16 commits, causing the recall of \texttt{commit} dropped from 62.89\% to 59.93\%. 
It is because the commits with similar patch patterns but applied across different components, 
    sometimes appear to be unrelated, despite potentially contributing to the same vulnerability remediation.
With only the dependency-driven strategy, 
    the system is unable to capture dispersed fixes that share similar remediation logic but are not explicitly dependent on one another.
Since partial sequence matching does not require comprehensiveness, 
    removing the coverage component does not adversely affect the overall performance.

\noindent
\textbf{Dependency-Driven Patch Reconstruction.}
Removing \textit{Dependency-Driven} leds to a more substantial performance degradation, 
    where a large portion of fixing commits could not be located, causing a decrease in \texttt{CVE-E}, i.e., precision dropped from 63.80\% to 41.72\%, recall dropped from 62.89\% to 50.83\%, and F1 dropped from 71.23\% to 53.13\%.
It is because \sysname can only identify commits independently through semantic analysis without dependency information. 
Eventually,
    \sysname may miss fixing commits with vague descriptions and misclassify functionality-related commits as vulnerability fixes when their descriptions are similar to those of anchor commits.


\noindent
\textbf{Patch Refinement.}
Removing patch refinement leds to noticeable degradation in precision for commit identification (i.e., \texttt{Commit})
    from 93.13\% to 69.9\%, while recall remains unchanged.
It is because candidate commits that are related to vulnerabilities, 
    but do not contribute to vulnerability remediation, 
    cannot be excluded effectively. 
 The absence of such refinement also negatively affected the performance of \texttt{CVE-E} and \texttt{CVE-P}, 
    causing the precision of \texttt{CVE-E} dropped to 39.88\% and  \texttt{CVE-P} to 60.12\%,

\vspace{0.05cm}
\noindent
\textbf{Impact of Different LLMs.}
As \sysname relies on LLMs for anchor commit identification and candidate refinement,
    we evaluated the impact of different LLMs on the overall performance of \sysname.
Specifically, 
    we replaced the default \texttt{gpt-5-nano} model used in \sysname with two alternative models: 
    \texttt{gpt-4o-mini}, which is smaller than the default model, and \texttt{gemini-3.1-flash}, which adopts a different model architecture.
To verify the necessity of \sysname, 
    rather than directly using LLMs for patch recognition,
    we further evaluated LLMs under a direct prompting setting for patch identification.

Table~\ref{tab:component} shows the result.
Both alternative models achieved performance comparable to the default model (gpt-5-nano, 75.08\%), yielding \texttt{Commit}-level F1 scores of 73.90\%~(\texttt{gpt-4o-mini}) and 73.83\%~(\texttt{gemini-3.1-flash}). 
For \texttt{CVE-E}, \texttt{gpt-4o-mini} and \texttt{gemini-3.1-flash} achieve comparable F1 scores of 71.20\% and 71.67\%, respectively, closely matching the 71.23\% achieved by \texttt{gpt-5-nano}.
For \texttt{CVE-P}, \texttt{gpt-4o-mini} and \texttt{gemini-3.1-flash} achieve comparable F1 scores of 81.76\% and 81.01\%, approaching the 84.66\% achieved by \texttt{gpt-5-nano}.
Furthermore, 
    when applying \texttt{gpt-5-nano} directly to patch identification without \sysname, the model failed to identify any complete multi-commit patch and produced no correct fixing commits.
These indicate that the superior performance of \sysname does not heavily rely on LLM capability.

\begin{figure}[t]
  \centering
  \includegraphics[width=0.8\linewidth]{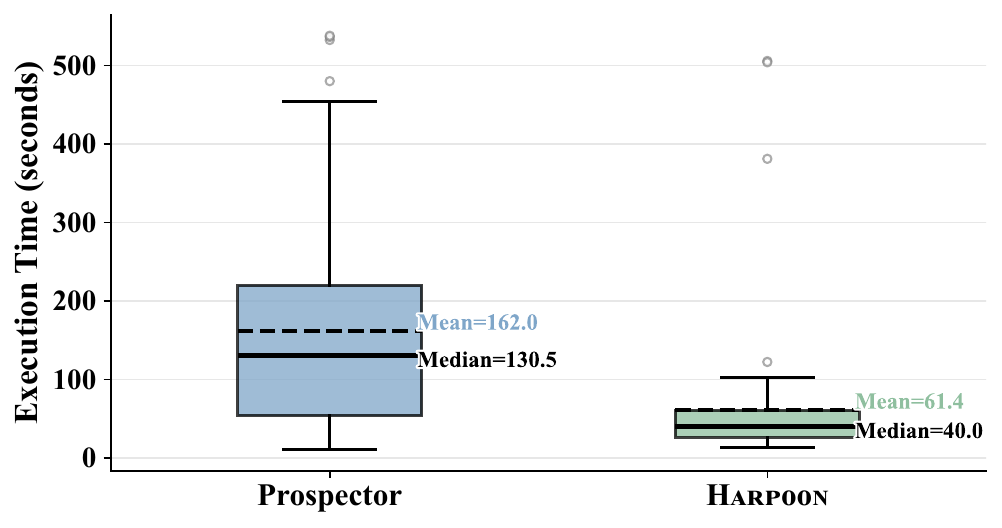}
  \caption{
  Execution Time Comparison.
  }
  \label{fig:exec_time}
\end{figure}

\subsection{RQ3: Computational Cost}
As \tool{Prospector} performed best among the state-of-the-art approaches, 
    we focused the computational-cost comparison on \sysname and \tool{Prospector}.
To avoid skewing the results due to failed or excessively time-consuming cases, 
    we conducted the cost evaluation only on cases successfully handled by both approaches.
The cost evaluation includes 828 vulnerabilities, consisting of 752 vulnerabilities with single-commit patches and 76 vulnerabilities with multi-commit patches. 
Overall, these vulnerabilities involve 934 fixing commits.
The comparison is shown in Figure~\ref{fig:exec_time}. 
Overall,
    \sysname achieved a mean execution time of 61.4s, substantially lower than the 162.0s required by \tool{Prospector}.
Similarly, the median execution time of \sysname is 40.0s, compared with 130.5s for \tool{Prospector},
    indicating that the improvement is not driven by a small number of favorable cases but is consistent across the benchmark. 
As illustrated by the boxplots, 
    \sysname exhibits a noticeably more consistent execution-time distribution, with a lower median, narrower interquartile range, and shorter whiskers than \tool{Prospector}.
Although there are two cases requiring more than 200s , 
    they correspond to uncommon cases that require substantially larger candidate spaces during patch analysis.
Overall, \sysname achieves both lower average cost and more stable execution behavior than \tool{Prospector}.

\subsection{RQ4: In-the-Wild Evaluation}

We further applied \sysname to a large-scale dataset, containing 30 projects with 5,146 CVEs. 
Due to the page limit, 
    the top-5 projects containing the largest number of multi-commit patches are reported in Table~\ref{tab:wild_patch_statistics}, 
    namely, 
    ImageMagick (104), FFmpeg (87), Wireshark (80), Apache HTTP Server (72), and Django (71). 
We found that these projects are typically large and long-maintained open-source systems with active maintenance across multiple branches and versions,
    making vulnerability remediation more likely to be distributed across multiple commits.
The rest results are available in our anonymized artifact~\cite{artifact}.
In total, 
    \sysname successfully identified 3,269 vulnerability patches, 
    including 2,513 single-commit patches and 756 multi-commit patches. 
All these analyses were completed within 68 hours.

\begin{table}[t]
\centering
\caption{Large-Scale Patch Identification Results across Representative Projects.}
\label{tab:wild_patch_statistics}
\resizebox{\linewidth}{!}{
\begin{tabular}{lrrrrr}
\toprule[1.5pt]
\textbf{Project} &
\textbf{\#Patches} &
\textbf{\#Multi} &
\textbf{Multi} &
\textbf{High} &
\textbf{Crit.} \\
\midrule
ImageMagick & 398   & 104 & 26.1\% & 23.1\% & 5.8\%  \\
FFmpeg      & 316   & 87  & 27.5\% & 40.2\% & 6.9\%  \\
Wireshark   & 418   & 80  & 19.1\% & 36.2\% & 0.0\%  \\
httpd       & 113   & 72  & 63.7\% & 38.9\% & 15.3\% \\
Django      & 98    & 71  & 72.4\% & 46.5\% & 12.7\% \\
\midrule
Top-5 Total & 1,343 & 414 & 30.8\% & 36.0\% & 7.7\%  \\
Overall     & 3,269 & 756 & 23.1\% & 36.5\% & 8.9\%  \\
\bottomrule[1.5pt]
\end{tabular}
}
\end{table}

While inspecting the results, 
    we found that vulnerabilities with multi-commit patches include 276 (36.5\%) \texttt{HIGH}-severity vulnerabilities and 67 (8.9\%) \texttt{CRITICAL}-severity vulnerabilities.
Most multi-commit patches involve only a few additional fixing commits, with two to four commits being the most common, accounting for 79.4\%.
There are also 9 vulnerabilities requiring more than ten commits for completed mitigation.
The largest multi-commit patch even contains 19 fixing commits, illustrating that comprehensive patch reconstruction can become more challenging for long remediation chains.
When examining the vulnerability remediation period, 
    the median remediation span is 4.2 days.
Approximately 48\% of multi-commit patches are completed within one week and 81\% within one month,
    although some remain unresolved for several months.
This highlights that vulnerability remediation may remain incomplete for extended periods, emphasizing the importance of timely identifying complete multi-commit patches before vulnerabilities are mistakenly considered fully remediated and deployed in practice.



\section{Discussion} \label{sec:discussion} 

\noindent\textbf{LLM Robustness.} \sysname relies on LLMs for semantic refinement and noise filtering of candidate multi-commit patches. Although LLMs are effective at distinguishing vulnerability-related commits from irrelevant noise, their decisions remain susceptible to hallucinations and reasoning errors. Consequently, incorrect commit selections may still occur, posing a potential threat to the robustness of \sysname.

\noindent \textbf{Static Analysis Reliability.} The dependency-driven reconstruction module leverages \tool{tree-sitter}–based static analysis to capture inter-commit dependency relationships. While efficient, this approach may be less reliable in complex codebases, particularly in the presence of implicit dependencies, potentially resulting in incomplete multi-commit patch reconstruction. Nevertheless, given a correctly identified anchor commit, the approach remains effective for the majority of cases.

%% file: Tables/rq1_srp.tex

\begin{table}[t]
\centering
\caption{Performance on Single-Commit Patch Identification.}
\begin{tabular}{c  c  c  c  c c}
\toprule[1.5pt]
\textbf{Tool} & \textbf{\#D} & \textbf{\#C} & \textbf{P(\%)} & \textbf{R(\%)} & \textbf{F1(\%)} \\
\midrule

\tool{VFCFinder}
& 278 & 76 & 27.34 & 9.30 & 13.89 \\

\tool{Prospector}
& 948 & 848 & 89.45 & 92.27 & 90.84 \\

\tool{SHIP}
& 917 & 603 & 65.76 & 85.53 & 74.35 \\

\tool{Taper}
& 799 & 0 & 0.00 & 0.00 & 0.00 \\

\midrule

\textbf{\sysname}
& 954 & 886 & \textbf{\underline{92.87}}
& \textbf{\underline{93.17}}
& \textbf{\underline{93.02}} \\

\bottomrule[1.5pt]
\end{tabular}
\begin{minipage}{\linewidth}
\footnotesize
\vspace{0.05cm}
\textit{Note:} \#D and \#C denote the numbers of detected and confirmed cases; P, R, and F1 denote Precision, Recall, and F1 Score.
\end{minipage}
\label{tab:rq1_srp}
\end{table}

%% file: Tables/rq1_irp_results.tex
\begin{table*}[t]
\centering
\caption{Performance on Multi-Commit Patch Identification.}
\label{tab:irp_benchmark}
\resizebox{\linewidth}{!}{
\begin{tabular}{c|ccccc|ccccc|ccccc}
\toprule[1.5pt]

\multirow{2}{*}{\textbf{Tool}}
& \multicolumn{5}{c|}{\textbf{Commit}}
& \multicolumn{5}{c|}{\textbf{CVE-E}}
& \multicolumn{5}{c}{\textbf{CVE-P}} \\

\cmidrule(lr){2-6}
\cmidrule(lr){7-11}
\cmidrule(lr){12-16}

& \textbf{\#D} & \textbf{\#C} & \textbf{P(\%)} & \textbf{R(\%)} & \textbf{F1(\%)}
& \textbf{\#D} & \textbf{\#C} & \textbf{P(\%)} & \textbf{R(\%)} & \textbf{F1(\%)}
& \textbf{\#D} & \textbf{\#C} & \textbf{P(\%)} & \textbf{R(\%)} & \textbf{F1(\%)} \\
\midrule

\tool{VFCFinder}$_{\textsc{Hint}}$
& 258 & 51 & 19.77 & 14.29 & 16.59
& 116 & 10 & 8.62 & 12.20 & 10.10
& 116 & 24 & 20.69 & 25.00 & 22.64 \\

\tool{VFCFinder}$_{\textsc{Orig}}$
& 341 & 36 & 10.56 & 6.39 & 7.96
& 130 & 7 & 5.38 & 3.72 & 4.40
& 130 & 7 & 5.38 & 3.72 & 4.40 \\

\tool{Prospector}$_{\textsc{Hint}}$
& 287 & 215 & 74.91 & 43.70 & 55.21
& 136 & 57 & 41.91 & 52.29 & 46.53
& 136 & 88 & 64.71 & 62.86 & 63.77 \\

\tool{Prospector}$_{\textsc{Orig}}$
& 470 & 285 & 60.64 & 50.62 & 55.18
& 161 & 84 & 52.17 & 44.68 & 48.14
& 161 & 84 & 52.17 & 44.68 & 48.14 \\

\tool{SHIP}
& 132 & 77 & 58.33 & 13.68 & 22.16
& 93  & 21 & 22.58 & 18.10 & 20.10
& 93  & 38 & 40.86 & 28.57 & 33.63 \\

\tool{Taper}
& 97,916 & 62 & 0.06 & 11.01 & 0.13
& 151 & 0 & 0.00 & 0.00 & 0.00
& 151 & 0 & 0.00 & 0.00 & 0.00 \\

\midrule

\textbf{\sysname}
&  364 & 339 & \textbf{\underline{93.13}} & \textbf{\underline{62.89}} & \textbf{\underline{75.08}}
& 163 & 104 & \textbf{\underline{63.80}} & \textbf{\underline{80.62}} & \textbf{\underline{71.23}}
& 163 & 138 & \textbf{\underline{84.66}} & \textbf{\underline{84.66}} & \textbf{\underline{84.66}} \\

\bottomrule[1.5pt]
\end{tabular}
}
\begin{minipage}{\textwidth}
\footnotesize
\vspace{0.05cm}
\textit{Note:} \#D and \#C denote the numbers of detected and confirmed cases, respectively; P, R, and F1 denote Precision, Recall, and F1 Score.
\end{minipage}
\end{table*}

%% file: Tables/rq2_ablation_study.tex
\begin{table*}[t]
\centering
\caption{Component Effectiveness Analysis of \sysname}

\resizebox{\linewidth}{!}{

\begin{tabular}{c | c  c  c  c  c | c  c  c  c  c | c  c  c  c  c}
\toprule[1.5pt]
\multirow{2}{*}{\textbf{Variant}} 
& \multicolumn{5}{c |}{\textbf{Commit}}                                  
& \multicolumn{5}{c |}{\textbf{CVE-E}}                             
& \multicolumn{5}{c}{\textbf{CVE-P}}                        \\ 
\cmidrule(lr){2-6}
\cmidrule(lr){7-11}
\cmidrule(lr){12-16}
& \textbf{\#D} & \textbf{\#C} & \textbf{P(\%)} & \textbf{R(\%)} & \textbf{F1(\%)} 
& \textbf{\#D} & \textbf{\#C} & \textbf{P(\%)} & \textbf{R(\%)} & \textbf{F1(\%)} 
& \textbf{\#D} & \textbf{\#C} & \textbf{P(\%)} & \textbf{R(\%)} & \multicolumn{1}{c}{\textbf{F1(\%)}} \\ 
\midrule

w/o \textit{Coverage}   
& 348 & 323  & 92.82 & 59.93 & 72.84 
& 163 & 88 & 53.99 & 77.88 & 63.77 
& 163 &  138 & 84.66 & 84.66 & 84.66 \\

w/o \textit{Dependency}      
& 298 & 274 & 91.95 & 50.83 & 65.47 
& 163 & 68 & 41.72 & 73.12 & 53.13 
& 163 & 138 & 84.66 & 84.66 & 84.66 \\

w/o \textit{Refinement}      
& 485 & 339 & 69.90  & \underline{64.57} & 67.13  
& 163 & 65 & 39.88 & 72.22 & 51.38 
& 163 & 98 & 60.12 & 79.67 & 68.53 \\

\midrule

w/ \textit{gpt-4o-mini}      
& 359 & 331 & 92.20  & 61.75 & 73.90  
& 155 & 104 & \underline{67.10} & 75.91 & 71.20 
& 158 & 130 & 82.28 & 81.25 & 81.76 \\

w/ \textit{gemini-3.1-flash}      
& 358 & 330 & 92.18  & 61.57 & 73.83  
& 155 & 105 & 67.74 & 76.09 & \underline{71.67} 
& 155 & 128 & 82.58 & 79.50 & 81.01 \\ 

\midrule

\textbf{\sysname}    
& \textbf{364} &  \textbf{\underline{339}} &  \textbf{\underline{93.13}} & \textbf{62.89}  & \textbf{\underline{75.08}}   
& \textbf{163} & \textbf{\underline{104}} & \textbf{63.80} & \textbf{\underline{80.62}} & \textbf{71.23}  
& 163 & \textbf{\underline{138}} & \textbf{\underline{84.66}} & \textbf{\underline{84.66}} & \textbf{\underline{84.66}}  \\

\bottomrule[1.5pt]
\end{tabular}
}
\begin{minipage}{\textwidth}
\footnotesize
\vspace{0.05cm}
\textit{Note:} \#D and \#C denote the numbers of detected and confirmed cases, respectively; P, R, and F1 denote Precision, Recall, and F1 Score.
\end{minipage}
\label{tab:component}
\end{table*}

%% file: Tex/relatedworks.tex

Existing research on vulnerability patch identification mainly falls into three categories: vulnerability-specific patch identification, silent fix identification, and empirical studies of vulnerability fixing practices. The first two develop fix identification techniques, whereas the third reveals the complexity of real-world vulnerability remediation. We briefly review each line of work below.

\vspace{0.05cm}
\noindent
\textbf{Vulnerability-specific patch identification.}
This line of research aims to identify patches for specific vulnerabilities.
Existing approaches~\cite{vcmatch,patchscout,patchfinder,promvpat,vfcfinder,prospector,patchfinder} typically model this task as one-to-one matching, 
    potentially assuming a single fixing commit per vulnerability, thereby overlooking multi-commit patches and leading to incomplete results.
The prior work, 
    SHIP~\cite{ship}, 
    relies on similarity-based filtering, which fails to capture multi-commit patch characteristics and thus cannot achieve comprehensive identification.

\vspace{0.05cm}
\noindent
\textbf{Silent fix identification.}
Unlike patch identification for known vulnerabilities, 
    this line of work focuses on \textbf{uncovering} vulnerability-fixing commits associated with undisclosed vulnerabilities from code changes and assessing their severity as an early warning mechanism.
Existing approaches primarily rely on deep learning~\cite{zhou2021finding,zhou2023colefunda,sun2023silent,wang2023graphspd,wen2025repository} and LLM-based reasoning~\cite{yang2025code}.
    \tool{EarlyVulnFix}~\cite{wen2024silent} leverages static analysis to identify vulnerability-fixing commits by analyzing code differences across versions.
Dong et al.~\cite{dong2025lies} further conduct an empirical study on the distribution, 
    silent window, and associated risks of silent vulnerability fixes in real-world projects.

\vspace{0.05cm}
\noindent
\textbf{Analysis of vulnerability fixing practices.}
Recent studies suggest that vulnerability fixing is more complex than commonly assumed.
Sun et al.~\cite{sun2023aspect} show that vulnerability databases often provide incomplete and inconsistent patch information, hindering reliable patch linkage.
Xie et al.~\cite{xie2024unveiling} demonstrate that vulnerability remediation frequently spans multiple commits rather than a single atomic fix.
Meanwhile, 
    Tan et al.~\cite{tan2022understanding} and Li et al.~\cite{li2024investigation} reveal the prevalence of cross-branch backporting, introducing additional delays and inconsistencies.
These findings collectively indicate that vulnerability fixing is an evolving, 
    multi-stage process, calling for process-aware modeling beyond simplified assumptions.

%% file: Tex/conclusion.tex

This paper addresses the underexplored problem of multi-commit patch identification.
We propose \sysname, 
    a unified framework integrating anchor commit identification, multi-commit patch reconstruction, and noise refinement.
Experiments show that \sysname significantly outperforms existing methods on both single- and multi-commit patch identification,
    achieving an F1 of 87.25\% at the commit level and over 90\% at the CVE level, with improvements exceeding 20\% compared with existing work.
In the wild,
    \sysname identifies 3,269 vulnerability patches, including 323 high-severity (CRITICAL/HIGH) cases previously absent from major databases (e.g., NVD, Debian Security, GHSA), covering 144 multi-commit patches.
We believe our approach and findings provide valuable insights into vulnerability remediation behaviors and can facilitate future research on patch identification.